\documentclass[twocolumn]{article}
\usepackage[utf8]{inputenc}
\usepackage{amsmath}
\usepackage{amsfonts}
\usepackage{listings}
\usepackage{hyperref}
\usepackage{graphicx}
\usepackage{float}
\usepackage{natbib}
\usepackage{multirow}
\usepackage{subcaption}
\usepackage[inline]{enumitem}
\usepackage{rotating}
\usepackage{float}
\usepackage{authblk}
\usepackage{amssymb}
\usepackage{color}
\usepackage{xspace}
\usepackage[table,xcdraw]{xcolor}
\usepackage[normalem]{ulem}
\usepackage{bm}
\usepackage{algorithm}
\usepackage{algorithmicx}
\usepackage{algpseudocode}

\makeatletter
\DeclareRobustCommand\onedot{\futurelet\@let@token\@onedot}
\def\@onedot{\ifx\@let@token.\else.\null\fi\xspace}
\def\etal{~et~al\onedot}
\def\eg{e.g\onedot} 
\def\ie{i.e\onedot}

\def \orop  {\ensuremath{\vee}}

\makeatother

\newfloat{lstfloat}{htbp}{lop}
\floatname{lstfloat}{Listing}

\definecolor{codegreen}{rgb}{0,0.6,0}
\definecolor{codegray}{rgb}{0.5,0.5,0.5}
\definecolor{codepurple}{rgb}{0.78,0,0.82}
\definecolor{backcolour}{rgb}{0.95,0.95,0.95}
\lstdefinestyle{mystyle}{
    backgroundcolor=\color{backcolour},   
    commentstyle=\color{codegreen},
    keywordstyle=\color{purple},
    numberstyle=\tiny\color{codegray},
    stringstyle=\color{codepurple},
    basicstyle=\ttfamily\footnotesize,
    breakatwhitespace=false,         
    breaklines=true,                 
    captionpos=b,                    
    keepspaces=true,                 
    numbers=left,                    
    numbersep=5pt,                  
    showspaces=false,                
    showstringspaces=false,
    showtabs=false,                  
    tabsize=2
}

\lstdefinelanguage{GLSL}%
{%
	morekeywords={%
		false,FALSE,NULL,true,TRUE,%
		__LINE__,__FILE__,__VERSION__,GL_core_profile,GL_es_profile,GL_compatibility_profile,%
		precision,highp,mediump,lowp,%
		break,case,continue,default,discard,do,else,for,if,return,switch,while,%
		void,bool,int,uint,float,double,vec2,vec3,vec4,dvec2,dvec3,dvec4,bvec2,bvec3,bvec4,ivec2,ivec3,ivec4,uvec2,uvec3,uvec4,mat2,mat3,mat4,mat2x2,mat2x3,mat2x4,mat3x2,mat3x3,mat3x4,mat4x2,mat4x3,mat4x4,dmat2,dmat3,dmat4,dmat2x2,dmat2x3,dmat2x4,dmat3x2,dmat3x3,dmat3x4,dmat4x2,dmat4x3,dmat4x4,sampler1D,sampler2D,sampler3D,image1D,image2D,image3D,samplerCube,imageCube,sampler2DRect,image2DRect,sampler1DArray,sampler2DArray,image1DArray,image2DArray,samplerBuffer,imageBuffer,sampler2DMS,image2DMS,sampler2DMSArray,image2DMSArray,samplerCubeArray,imageCubeArray,sampler1DShadow,sampler2DShadow,sampler2DRectShadow,sampler1DArrayShadow,sampler2DArrayShadow,samplerCubeShadow,samplerCubeArrayShadow,isampler1D,isampler2D,isampler3D,iimage1D,iimage2D,iimage3D,isamplerCube,iimageCube,isampler2DRect,iimage2DRect,isampler1DArray,isampler2DArray,iimage1DArray,iimage2DArray,isamplerBuffer,iimageBuffer,isampler2DMS,iimage2DMS,isampler2DMSArray,iimage2DMSArray,isamplerCubeArray,iimageCubeArray,atomic_uint,usampler1D,usampler2D,usampler3D,uimage1D,uimage2D,uimage3D,usamplerCube,uimageCube,usampler2DRect,uimage2DRect,usampler1DArray,usampler2DArray,uimage1DArray,uimage2DArray,usamplerBuffer,uimageBuffer,usampler2DMS,uimage2DMS,usampler2DMSArray,uimage2DMSArray,usamplerCubeArray,uimageCubeArray,struct,%
		gl_BackColor,gl_BackLightModelProduct,gl_BackLightProduct,gl_BackMaterial,gl_BackSecondaryColor,gl_ClipDistance,gl_ClipPlane,gl_ClipVertex,gl_Color,gl_DepthRange,gl_DepthRangeParameters,gl_EyePlaneQ,gl_EyePlaneR,gl_EyePlaneS,gl_EyePlaneT,gl_Fog,gl_FogCoord,gl_FogFragCoord,gl_FogParameters,gl_FragColor,gl_FragCoord,gl_FragData,gl_FragDepth,gl_FrontColor,gl_FrontFacing,gl_FrontLightModelProduct,gl_FrontLightProduct,gl_FrontMaterial,gl_FrontSecondaryColor,gl_InstanceID,gl_Layer,gl_LightModel,gl_LightModelParameters,gl_LightModelProducts,gl_LightProducts,gl_LightSource,gl_LightSourceParameters,gl_MaterialParameters,gl_ModelViewMatrix,gl_ModelViewMatrixInverse,gl_ModelViewMatrixInverseTranspose,gl_ModelViewMatrixTranspose,gl_ModelViewProjectionMatrix,gl_ModelViewProjectionMatrixInverse,gl_ModelViewProjectionMatrixInverseTranspose,gl_ModelViewProjectionMatrixTranspose,gl_MultiTexCoord0,gl_MultiTexCoord1,gl_MultiTexCoord2,gl_MultiTexCoord3,gl_MultiTexCoord4,gl_MultiTexCoord5,gl_MultiTexCoord6,gl_MultiTexCoord7,gl_Normal,gl_NormalMatrix,gl_NormalScale,gl_ObjectPlaneQ,gl_ObjectPlaneR,gl_ObjectPlaneS,gl_ObjectPlaneT,gl_Point,gl_PointCoord,gl_PointParameters,gl_PointSize,gl_Position,gl_PrimitiveIDIn,gl_ProjectionMatrix,gl_ProjectionMatrixInverse,gl_ProjectionMatrixInverseTranspose,gl_ProjectionMatrixTranspose,gl_SecondaryColor,gl_TexCoord,gl_TextureEnvColor,gl_TextureMatrix,gl_TextureMatrixInverse,gl_TextureMatrixInverseTranspose,gl_TextureMatrixTranspose,gl_Vertex,gl_VertexID,%
		gl_MaxClipPlanes,gl_MaxCombinedTextureImageUnits,gl_MaxDrawBuffers,gl_MaxFragmentUniformComponents,gl_MaxLights,gl_MaxTextureCoords,gl_MaxTextureImageUnits,gl_MaxTextureUnits,gl_MaxVaryingFloats,gl_MaxVertexAttribs,gl_MaxVertexTextureImageUnits,gl_MaxVertexUniformComponents,%
		abs,acos,all,any,asin,atan,ceil,clamp,cos,cross,degrees,dFdx,dFdy,distance,dot,equal,exp,exp2,faceforward,floor,fract,ftransform,fwidth,greaterThan,greaterThanEqual,inversesqrt,length,lessThan,lessThanEqual,log,log2,matrixCompMult,max,min,mix,mod,noise1,noise2,noise3,noise4,normalize,not,notEqual,outerProduct,pow,radians,reflect,refract,shadow1D,shadow1DLod,shadow1DProj,shadow1DProjLod,shadow2D,shadow2DLod,shadow2DProj,shadow2DProjLod,sign,sin,smoothstep,sqrt,step,tan,texture1D,texture1DLod,texture1DProj,texture1DProjLod,texture2D,texture2DLod,texture2DProj,texture2DProjLod,texture3D,texture3DLod,texture3DProj,texture3DProjLod,textureCube,textureCubeLod,transpose,%
		rgb
	},
	sensitive=true,%
	morecomment=[s]{/*}{*/},%
	morecomment=[l]//,%
	morestring=[b]",%
	morestring=[b]',%
	moredelim=*[directive]\#,%
	moredirectives={define,defined,elif,else,if,ifdef,endif,line,error,ifndef,include,pragma,undef,warning,extension,version}%
}[keywords,comments,strings,directives]%
\makeatletter
\let\@fnsymbol\@arabic
\makeatother

\title{Decoupled Boundary Handling in SPH}
\author{Rustam Akhunov\thanks{rustam.akhunov@uni-siegen.de, corresponding author}, Andreas Kolb\thanks{andreas.kolb@uni-siegen.de}}

\affil{University of Siegen, 57076 Siegen, Germany}

\date{}

\begin{document}

\maketitle

\begin{abstract}
Particle-based boundary representations are frequently used in \emph{Smoothed Particle Hydrodynamics (SPH)} due to their simple integration into fluid solvers.
Commonly, incompressible fluid solvers estimate the current density and corresponding forces in case the current density exceeds the rest density to push fluid particles apart. 
Close to the boundary, the calculation of the fluid particles' density involves both, neighboring fluid and neighboring boundary particles, yielding an overestimation of density, and, subsequently, wrong pressure forces and wrong velocities leading to the disturbed fluid particles' behavior in the vicinity of the boundary. 

In this paper, we present a detailed explanation of this disturbed fluid particle behavior, which is mainly due to the \emph{combined} or \emph{coupled} handling of the fluid-fluid particle and the fluid-boundary particle interaction. 
We propose the \emph{decoupled} handling of both interaction types, leading to two densities for a given fluid particle, i.e., fluid-induced density and boundary-induced density. In our approach, we alternately apply the corresponding fluid-induced and boundary-induced forces during pressure estimation. 
This separation avoids force overestimation and reduces unintended fluid dynamics near the boundary, as well as a consistent fluid-boundary distance across different fluid amounts and different particle-based boundary handling methods.
We compare our method with two regular state-of-the-art methods in different experiments and show how our method handles detailed boundary shapes.
\end{abstract}

\section{Introduction}
\label{s:intro}
Boundary handling is an important part of fluid simulation using \emph{Smoothed Particle Hydrodynamics (SPH)}. There are several approaches to handle the interaction of SPH-based fluids with boundaries, \ie, particle-based~\cite{akinci2012versatile}, external boundary handling methods~\cite{losasso2008two} or direct boundary integral methods~\cite{koschier2017density}.

In particle-based boundary handling, a particle's density is usually calculated by taking all neighboring particles into account, \ie, fluid and boundary particles. Whereas including fluid particles in the neighborhood is a regular rule in SPH, including static boundary particles causes the wrong estimation of forces between fluid particles in the vicinity of boundaries, as we demonstrate in this paper. This yields distances between the fluid and the boundary that depend on the number of fluid particles and result in inconsistent fluid behavior and  unintended fluid dynamics close to boundaries (see, for instance, Fig.~\ref{fig:teaser}). 

In this paper, we first analyze the cause of the inconsistent fluid behavior close to boundaries, and we propose a new method for \emph{decoupled boundary density calculation}, which divides the density calculation into the \emph{boundary-induced density} and the \emph{fluid-induced density}. While the first one takes only the neighboring boundary particles of a given fluid particle into account, the latter accounts for the interaction of a given fluid particle with its neighboring fluid particles. 

We also propose a \emph{decoupled pressure calculation method}, which alternates between the boundary-induced density and the fluid-induced density. This separation drastically reduces the overestimation of forces appearing in the standard coupled version and makes the fluid quantities more consistent across the complete fluid domain, e.g., it reduces residual oscillations caused by force overestimation.

In our experiments, we integrate our method with the IISPH solver~\cite{ihmsen2013implicit} (in the variant by Band\etal\citeyear{band2018pressure}) using two types of boundary pressure estimators, \ie, Pressure Mirroring (PM)~\cite{akinci2012versatile} and Pressure Boundaries (PB)~\cite{band2018pressure}. Comparing the decoupled with the mixed methods, the decoupled approach demonstrates significant benefits in terms of residual oscillations, residual pressure, and velocity, moreover, the decoupled method also brings more consistency across different pressure estimators.

Conceptually, our decoupled approach can be combined with any scheme for estimating the density of the boundary particles, and it can be applied to non-particle-based boundary handling schemes as well. 

\section{Prior Work}
\label{s:prior}
\begin{figure}[t]
  \centering
  \includegraphics[height=\linewidth]{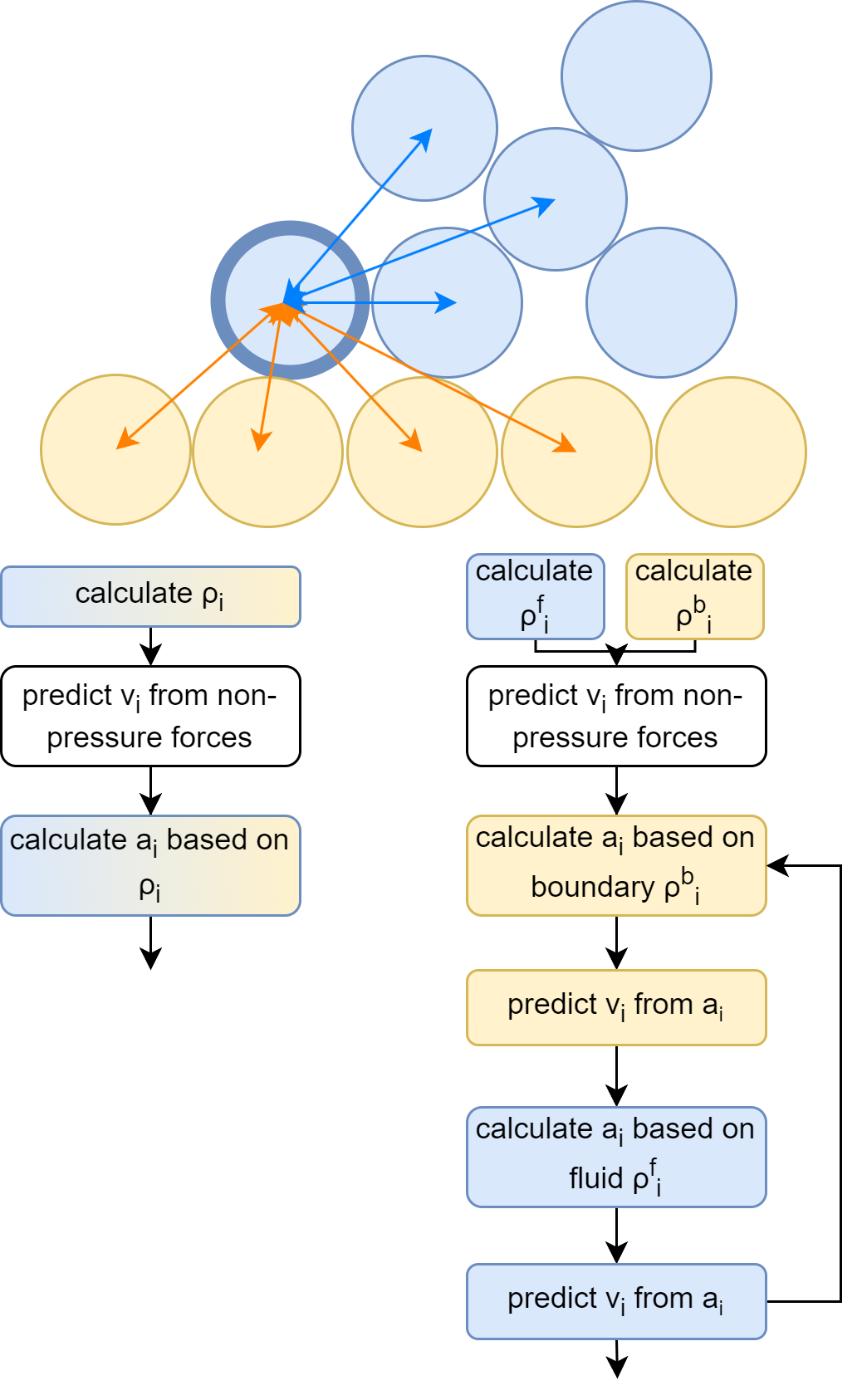}
  \caption{Depiction of the fluid and boundary neighbors of the selected (bold blue line) fluid particle. Arrows represent an interaction between the selected fluid particle and its neighboring fluid particles (blue) and boundary particles (yellow). Blue lines represent interaction with fluid neighbors, and yellow arrows represent interaction with boundary particles. The left flowchart is the classical approach in density and pressure calculation, and the right flowchart depicts our decoupled approach}
  \label{fluid-fluid-boundary}
\end{figure}

One of the first SPH approaches in fluid-boundary interaction is described by Monaghan~\cite{monaghan1994simulating}, where the boundary is represented by particles and the interaction between fluid and boundary particles is ruled by the Lennard-Jones potential.
Later, alternative methods of calculation of the forces between fluid and boundary particles have been proposed, such as direct forcing by Becker\etal\cite{becker2009direct} or pressure forces by Akinci\etal\citeyear{akinci2012versatile}.
Modern approaches in boundary handling estimate pressure values for the boundary particles that yield forces and resulting accelerations for fluid particles that prevent penetration of fluid particles into the boundary~\cite{koschier2020smoothed,koschier2022survey}.

Several approaches to pressure calculation for boundary particles are used in computer graphics. The Pressure Mirroring (PM) method~\cite{akinci2012versatile} estimates the pressure for a boundary particle by directly assigning the pressure of the currently interacting fluid particle. Other methods use a pressure extrapolation approach, where the boundary  particles' pressure is extrapolated based on the pressure and position of the neighboring fluid particles. Specifically, the Pressure Boundary (PB) method uses a separate pressure solver to estimate a unique pressure value for each boundary particle~\cite{band2018pressure}. An alternative approach uses a Moving Least Squares pressure extrapolation to calculate the pressure at boundary particles, where the pressure values of the boundary particles correspond to the pressure values of the fluid particles in its vicinity~\cite{band2018mls}.

Beyond particle-based boundary handling methods, other approaches have been proposed, such as direct forcing without boundary particles applied to fluid particles that approach the boundary surface, as done by Bodin\etal\citeyear{bodin2011constraint}. 
Other approaches to non-particle-based boundary representation include the meshed-based method by Huber\etal\citeyear{huber2015boundary}, the distance function-based method by Harada\etal\citeyear{harada2007smoothed}, the density maps by Bender\etal\citeyear{bender2019volume}, and the semi-analytical boundary handling by Winchenbach\etal\citeyear{winchenbach2020semi}.

Almost all of these boundary handling methods perform density calculations to deduce pressure forces that prevent boundary penetration. These methods estimate the density using both, boundary- and fluid-induced densities simultaneously in a \emph{coupled way}, leading to an overestimation of pressure and, hence, forces, and leads to the excess motion of the fluid particles; see detailed discussion in Sec.\ref{s:method}.

Our approach of decoupled density estimation borrows from a similar concept proposed by Gissler\etal\citeyear{gissler2019interlinked} applied in the context of boundary-boundary interactions, i.e., for rigid bodies collision reaction. For the calculation of regular SPH quantities, Gissler\etal\citeyear{gissler2019interlinked} use standard ``coupled'' density estimation, while they exclude fluid particles when estimating forces between rigid objects.

\section{Method}
\label{s:method}

\subsection{Problem Statement}
\label{s:method.problem}

The main challenge regarding boundary handling in SPH is the calculation of forces induced by the boundary on the fluid particles. The boundary must have at least the property of impenetrability, \ie, the forces induced by the boundary must be large enough to push the approaching fluid particles back to prevent penetration. 

Fig.~\ref{fluid-fluid-boundary} depicts the interaction of a single fluid particle (highlighted with a dark blue outline) with neighboring fluid (blue) and boundary (yellow) particles, where blue and orange arrows represent interactions with fluid and boundary neighbors, respectively. Both types of interaction serve different goals, that is, the interaction with the boundary particles in the fluid particle's neighborhood provides impenetrability, while the interaction with the neighboring fluid particles enables fluid dynamics.  

\begin{figure}[h]
  \centering
  \includegraphics[width=\linewidth]{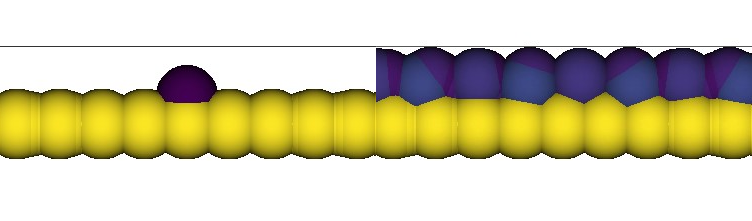}
  \caption{Comparison of the fluid-boundary distance for a single fluid particle (left) and a fluid sheet with several fluid neighbors. The boundary particles and fluid particles are colored yellow and blue-purple, respectively. The top gray line shows the height of the fluid on the right. Screenshot from our SPH simulation}
  \label{fb_distance}
  
\end{figure}

Commonly in SPH, the calculation of the density and the pressure forces for the given fluid particle considers all neighboring fluid and boundary particles in a \emph{coupled fashion} (see Fig.~\ref{fluid-fluid-boundary} left flow chart).
However, this coupling of boundary and fluid interaction causes an overestimation of SPH quantities for both interactions.
More precisely, the density and, subsequently, the repulsive pressure force between the given fluid particle and the boundary depends on the number of fluid neighbors in its proximity. Thus, an individual fluid particle without any fluid neighbors will have less density and will receive less repulsive pressure force than a fluid particle with some fluid neighborhood. In practice, this yields a varying distance between fluid particles and the boundary, where fluid particles with many fluid particle neighbors have a larger distance to the boundary surface than a fluid particle with few or no fluid particle neighbors (see Fig.~\ref{fb_distance}). Likewise, the calculation of fluid-fluid pressure or other forces, \eg, surface tension and artificial viscosity, within the boundary interaction radius are affected by the overestimation of the fluid particle's density due to boundary interaction, resulting in higher inter-fluid-particle forces, and, hence, in unintended fluid dynamics. 

\begin{algorithm}
\caption{The decoupling algorithm}\label{algo1:common}
\begin{algorithmic}[1]
\While{animating}
    \ForAll{particle i}
        \If{i fluid particle}
            \State calculate $\rho^b_i$ and $\rho^f_i$ 
        \Else
            \State set $\rho^b_i$ and $\rho^f_i$ equal to $\gamma$
        \EndIf
        \State initialize $a^{f}_i$ by $a_i$ \Comment{$a_i$: non-press. acc.}
    \EndFor
    \While{err$_b \ge$  lim$_b \orop$ err$_f \ge$ lim$_f \orop$ iter$_g \le$ iter$_{\text{max}}$}
        \While{err$_b \ge$ lim$_b \orop$ iter $\le$ iter$_{\text{max}}$}
        \State\Comment{bound. interact.}
            \ForAll{fluid particle i}
                    \State predict $v_i$ from $a^{f}_i$
                    \State calculate pressure $p_i^b$ using $\rho^b_i$
                    \State update $a^{b}_i$
                    \State calculate err$_b$
            \EndFor
        \EndWhile
        \While{err$_f \ge$ lim$_f \orop$ iter $\le$ iter$_{\text{max}}$} 
        \State\Comment{fluid interaction}
            \ForAll{fluid particle i}
                    \State predict $v_i$ from $a^{b}_i$ 
                    \State calculate pressure $p_i^f$ using $\rho^f_i$
                    \State update $a^{f}_i$
                    \State calculate err$_f$
            \EndFor
        \EndWhile
    \EndWhile
    \State update $v_i$, update positions
\EndWhile
\end{algorithmic}
\end{algorithm}

\subsection{Decoupled Boundary Handling}
\label{s:method.decoup}
In the following, we build upon the concept of \emph{artificial density} introduced by Akinci\etal\citeyear{akinci2013versatile}, which was later improved by Band\etal\citeyear{band2018pressure}. The concept of artificial density is additionally used to the regular density. Artificial density is dimensionless and represents the ratio between the current regular density $\rho$ and the rest density $\rho_0$, i.e., $\rho_i^a=\rho/\rho_0$. Artificial density values $\ge 1$, for example, result in an overcompression, yielding corresponding pressure forces. 

First, we reformulate the artificial density calculation by separating it into three components:
\begin{equation} \label{eq1:density}
    \rho_i = \sum_j V_j \cdot W_{ij} = V_i \cdot W_{ii} + \sum_f V_f \cdot W_{if} + \sum_b V_b \cdot W_{ib}
\end{equation}
where $i$ is the index of the current fluid particle, $j$ is the index for all neighboring particles, $f$ is the index for neighboring fluid particles, and $b$ is the index for neighboring boundary particles. $V_{i,j,f,b,}$ denote the rest volumes of the particles (similar to Band\etal\citeyear{band2018pressure}) and $W_{ij}$ is the kernel value between particles $i$ and $j$ (likewise for indices $f,b$). The first component in Eq.~\eqref{eq1:density} is the ``self-contribution'' of the particle, the second component is a fluid neighbors' contribution, and the third component is a boundary neighbors' contribution.

For our method, we derive two densities. 
The \emph{fluid induced density} for the fluid particle $i$ is calculated by omitting the boundary neighbor contribution
\begin{equation} \label{eq1:density1}
    \rho^f_i =  V_i \cdot W_{ii} + \sum_f V_f \cdot W_{if},
\end{equation}
and the \emph{boundary induced density} for the fluid particle $i$ is calculated by omitting the fluid neighbors contribution component
\begin{equation} \label{eq1:density2}
    \rho^b_i =  V_i \cdot W_{ii} + \sum_b V_b \cdot W_{ib}.
\end{equation}

Note, that $\rho_i \neq \rho^f_i + \rho^b_i$. The main idea of this separation is that $\rho^f_i$ represents the pure fluid-fluid interaction which is not affected by neighboring boundary particles, whereas $\rho^b_i$ represents the pure fluid-boundary interaction not affected by neighboring fluid particles.

Analogous to Band\etal\citeyear{band2018pressure}, we use a \emph{constant boundary density} $\rho^{bb}_i = \gamma$. However, as we use $\gamma$ only  for the boundary induced density without considering fluid particles, we use a slightly reduced boundary density of $\gamma=0.6$ instead of $\gamma=0.7$ as proposed by Band\etal\citeyear{band2018pressure}.

Finally, the given iterative pressure solver is split into two stages, which are executed alternately, as shown in Alg.~\ref{algo1:common}. The first stage comprises the calculation of the \emph{boundary pressure} induced by fluid-boundary interaction using $\rho^b_i$, while the second stage calculates the \emph{fluid pressure} induced by fluid-fluid interaction using $\rho^f_i$. For both stages, the error thresholds, lim$_b$ and lim$_f$, and the maximum number of iterations iter$_{max}$ are used to control convergence. These calculations are alternatively executed until both of the errors, err$_b$ and err$_f$, are less than their thresholds or the number of global iterations $iter_g$ is more than iter$_{max}$. 
Note, that the calculation of the current \emph{boundary pressure} $p_i^b$ requires the \emph{fluid induced} acceleration $a_i^f$, and, correspondingly, the calculation of the current \emph{fluid pressure} $p_i^f$ requires the \emph{boundary induced} acceleration $a_i^f$. The initial fluid acceleration is taken from the non-pressure forces. 
%

\begin{figure*}[h]
  \centering
  \includegraphics[width=\linewidth]{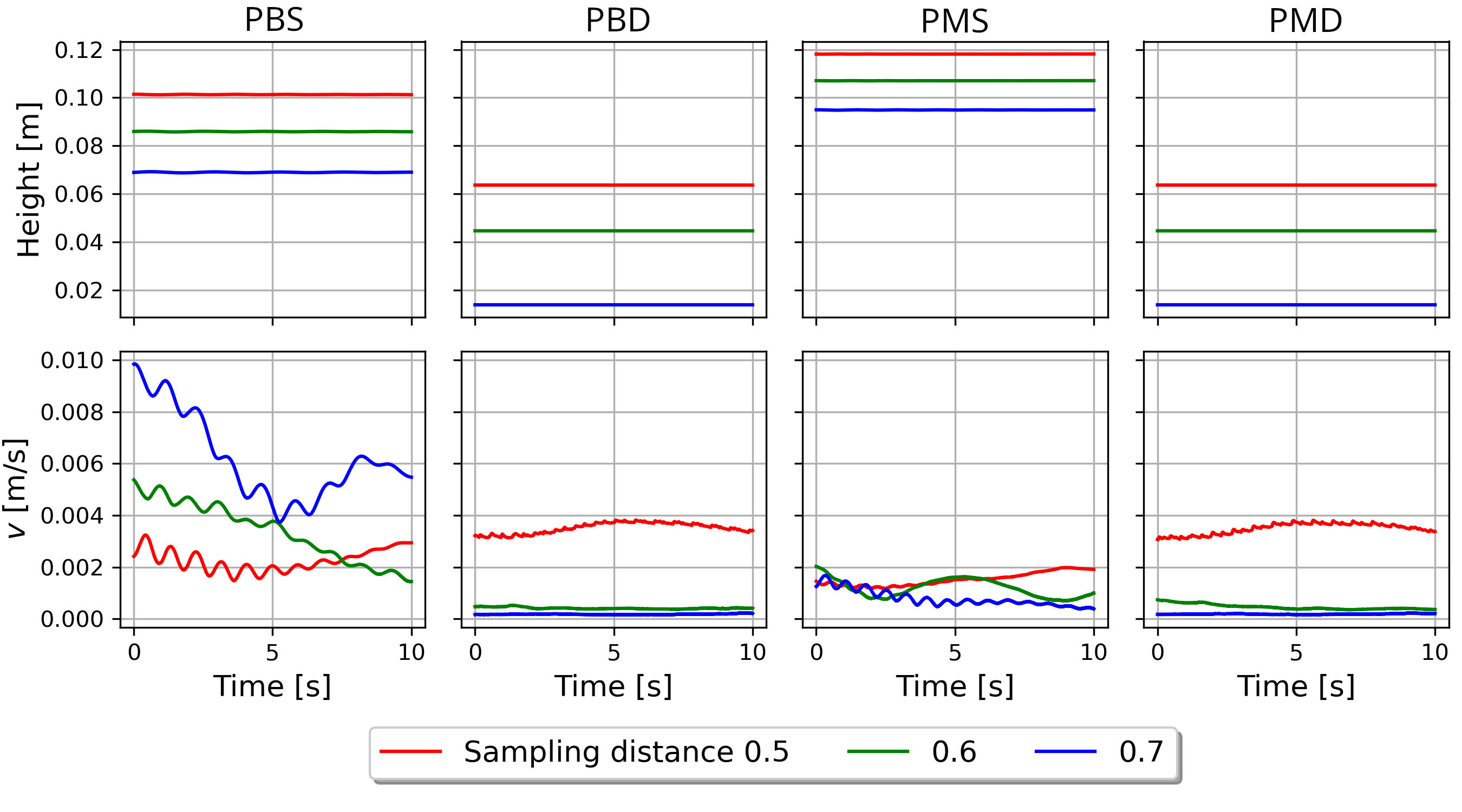}
  \caption{Resting fluid sheet experiment. The average height values over time (top) and the average velocity amplitude (bottom) are shown. The experiment is done for different sampling distances. The first $10$~s are shown after the relaxation criteria have been reached.}
  \label{fluid_sheet_relax}
\end{figure*}

\begin{figure*}
   \includegraphics[width=\textwidth]{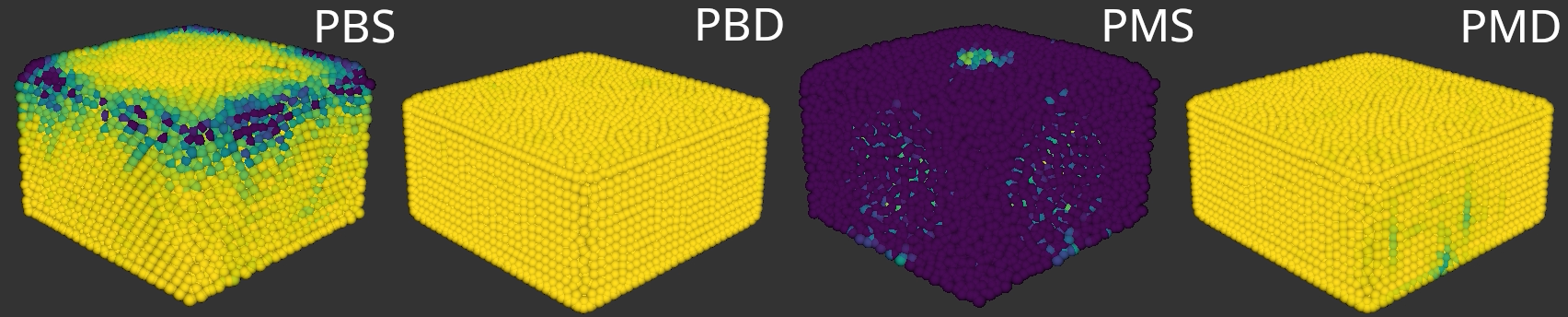}
   \caption{Resting fluid bulk experiment. Comparison of the pressure boundaries (PB$\ast$) and pressure mirroring (PM$\ast$) methods in their standard ($\ast$S) with the proposed decoupled ($\ast$D) implementations. The colormap encodes the velocity amplitude between $0$~$m/s$ (yellow) and $0.1$~$m/s$ (dark blue) $8$~s after the fluid's relaxation is completed.}
   \label{fig:teaser}
\end{figure*}
 
\begin{figure*}[h]
  \centering
  \includegraphics[width=400pt]{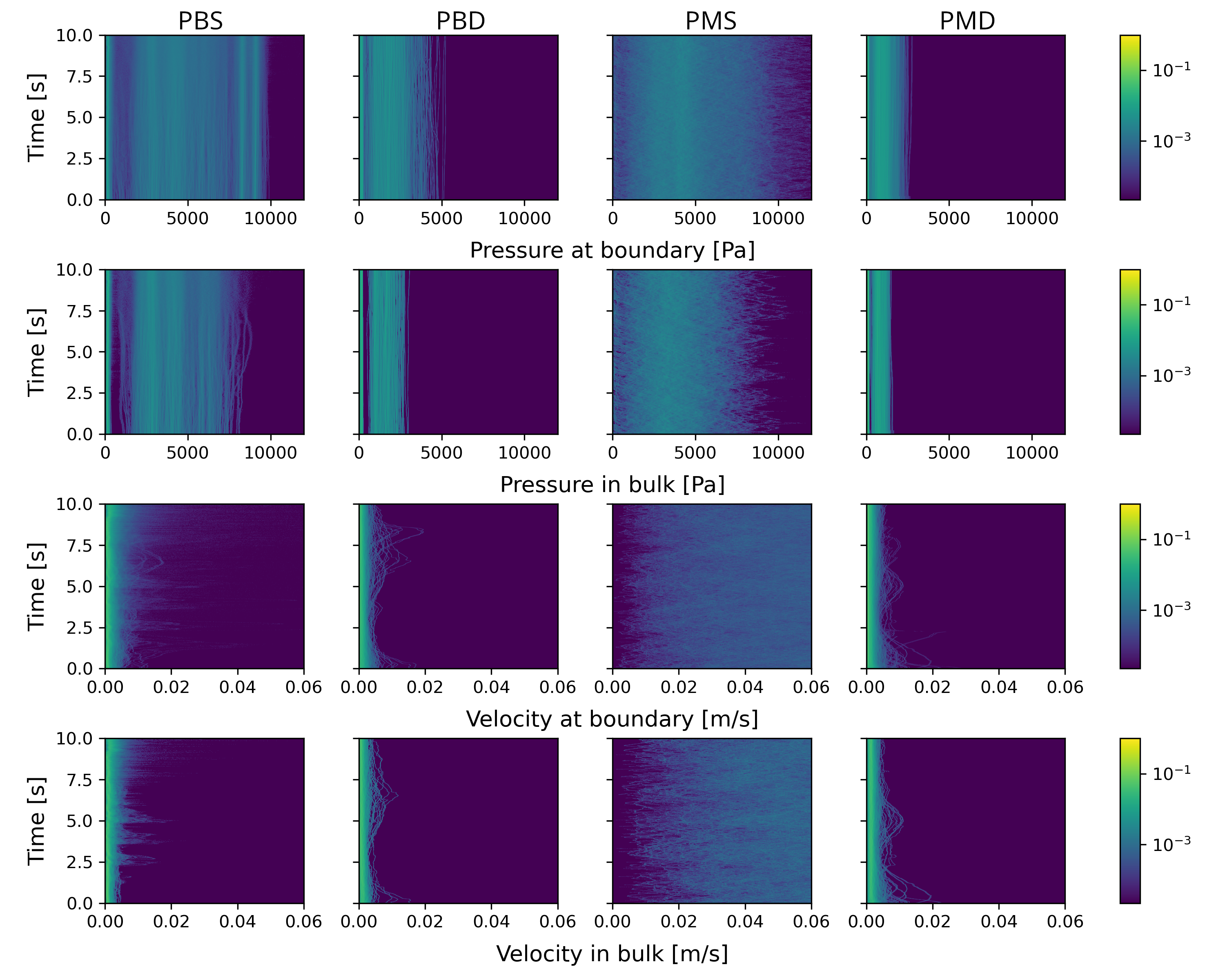}
  \caption{Resting fluid bulk experiment. The pressure and velocity histograms over time are presented. The measure of pressure and velocity is done for particles in bulk and for particles interacting with the boundary separately.}
  \label{still_water}
\end{figure*}

\begin{figure*}[h]
  \centering
  \includegraphics[width=\linewidth]{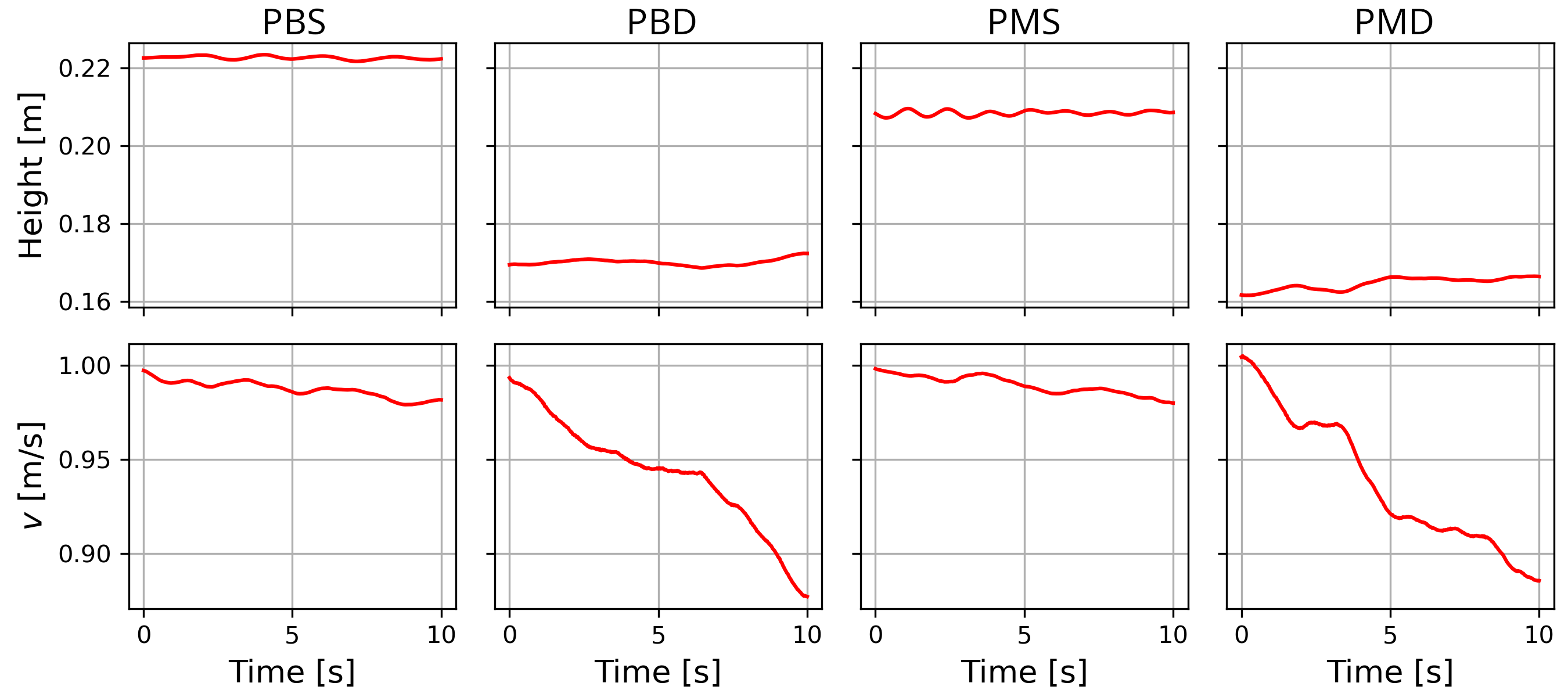}
  \caption{Fluid Sheet with tangential motion. The height values and the average velocity of the fluid particles during $8$~s  of sliding after the acceleration are shown.}
  \label{fluid_sheet_slide}
  
\end{figure*}

\begin{figure}[h]
  \centering
  \includegraphics[width=\linewidth]{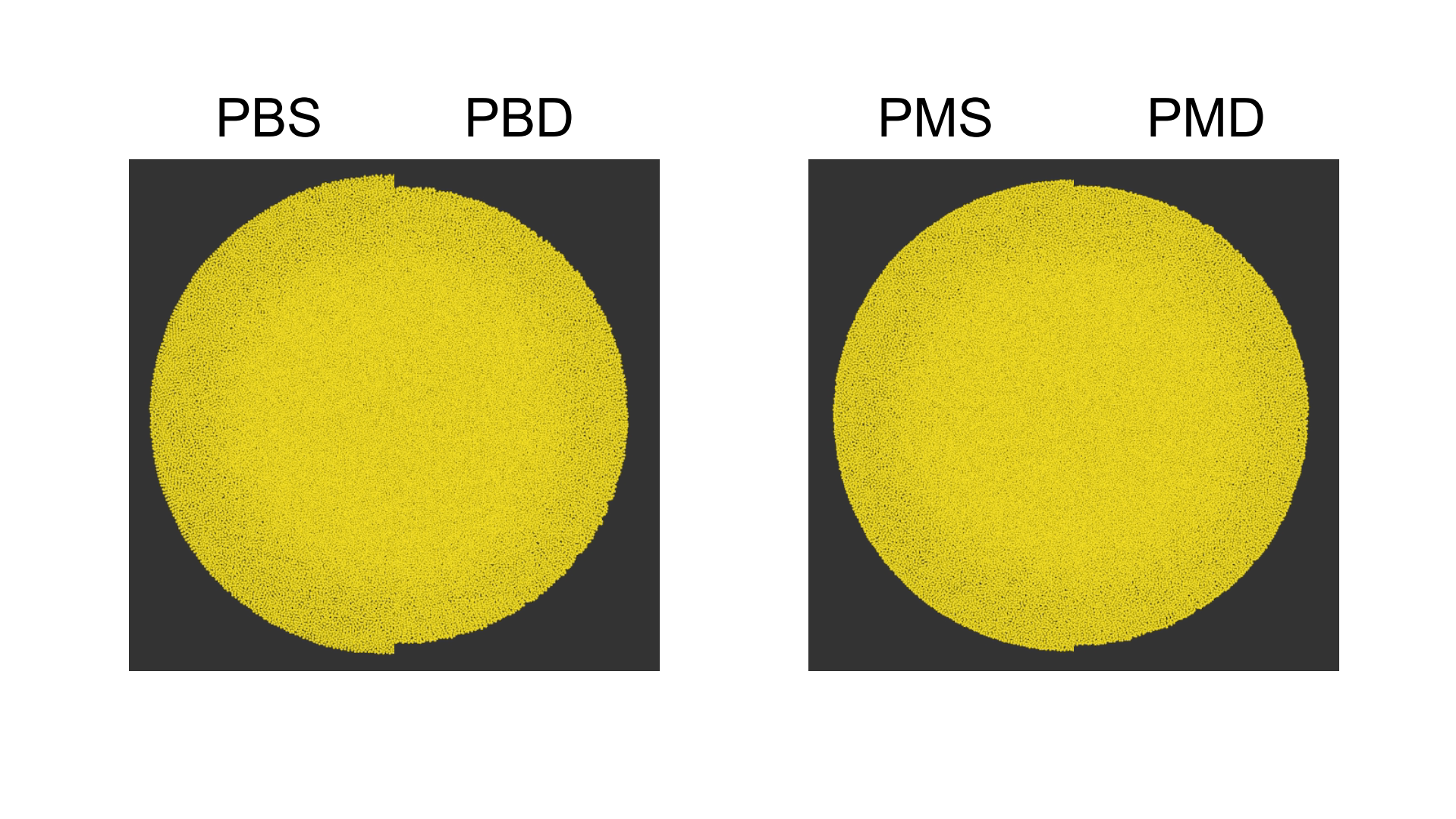}
  \caption{Cylinder experiment. Top-down view of the distribution of the particles $1$~s after cylinder release.}
  \label{cylinder}
\end{figure}

\begin{figure}[h]
  \centering
  \includegraphics[width=\linewidth]{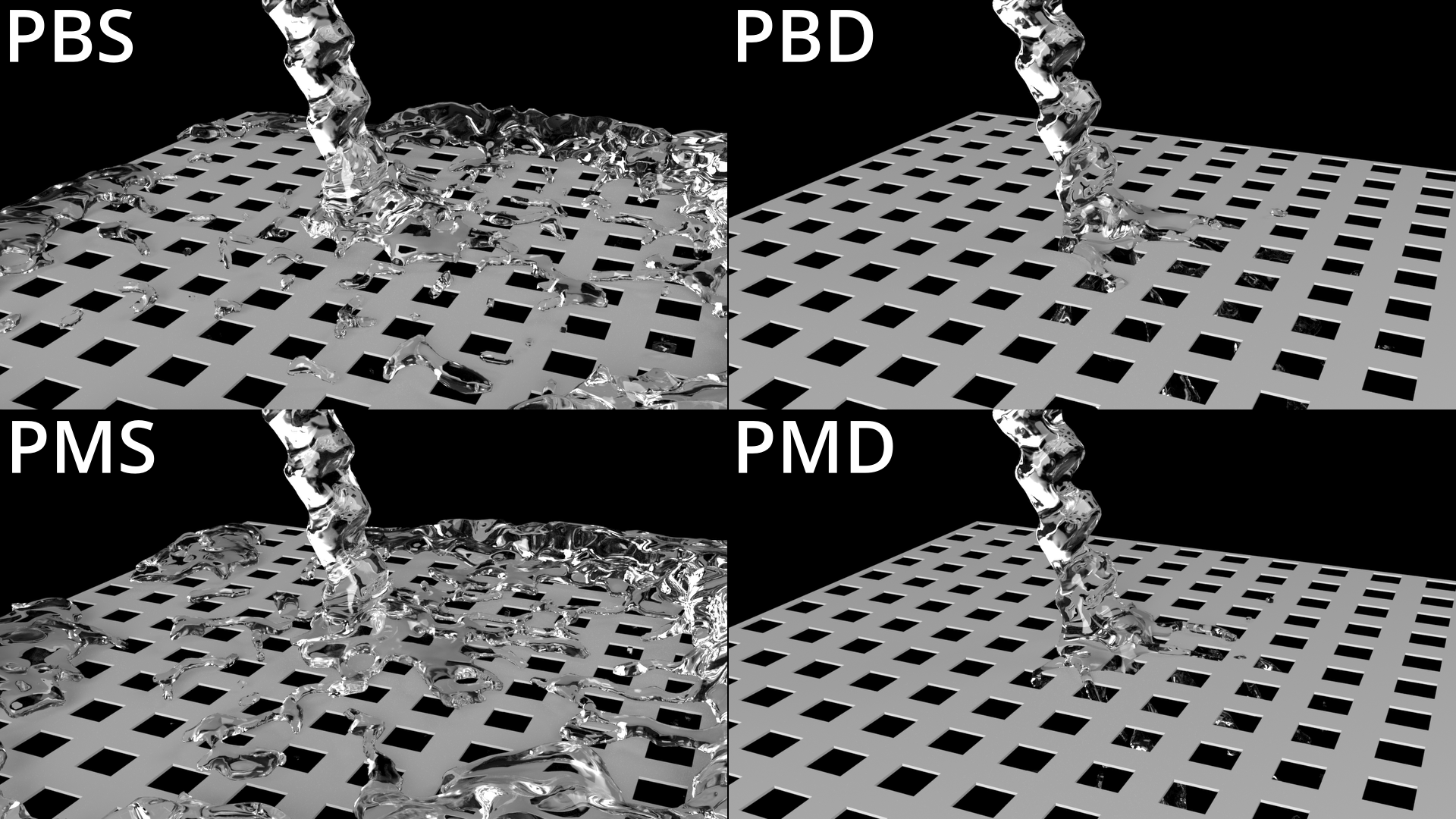}
  \caption{Water jet passing a grid experiment. The close-up of the jet going through the grid.}
  \label{fig:container_holes2}
  
\end{figure}

\begin{figure*}[h]
  \centering
  \includegraphics[width=\linewidth]{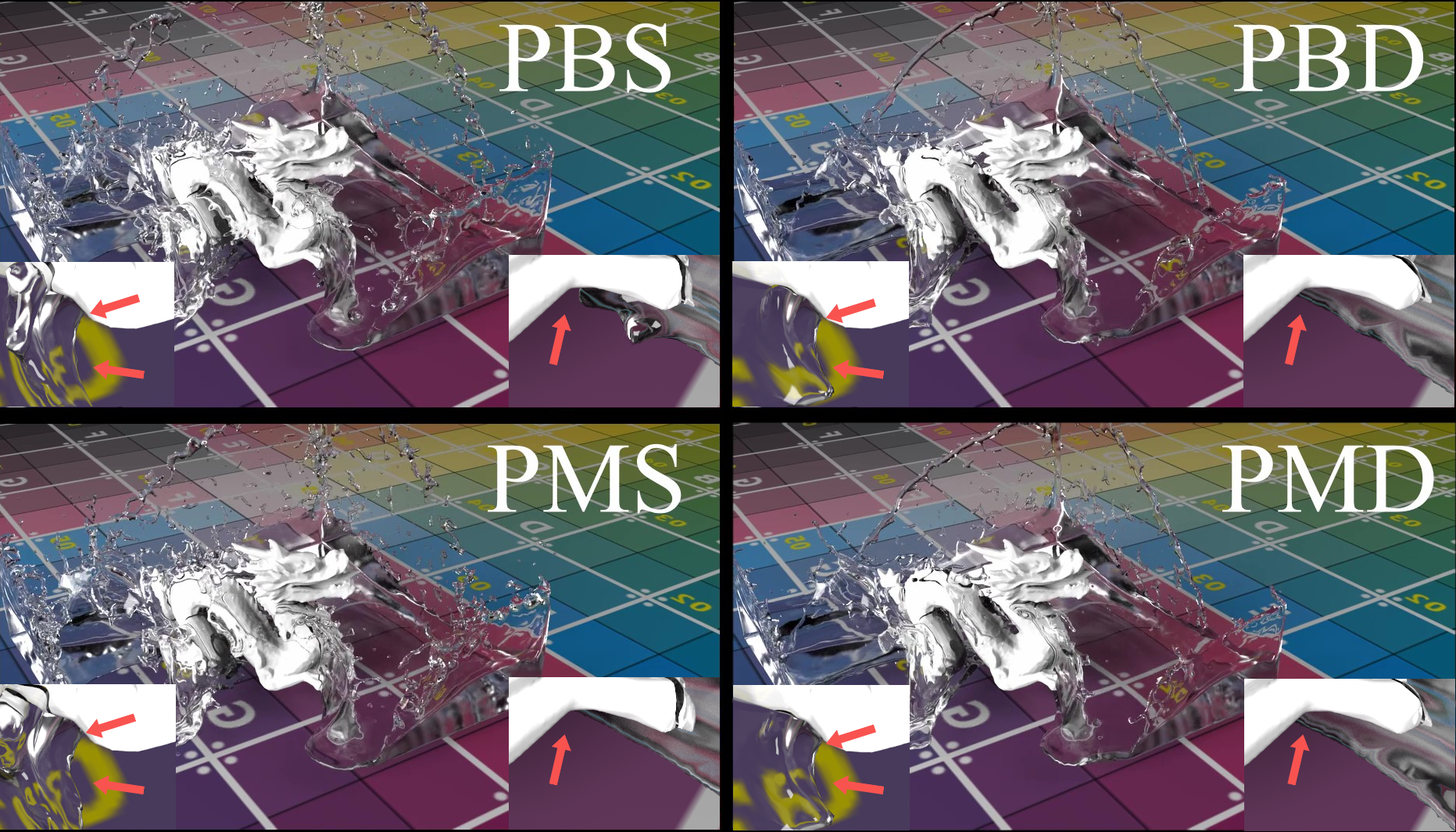}
  \caption{Dragon dam break experiment. The state of the scene at $2$s simulation time. The left bottom corner is a close-up of the dragon's tail, and the right bottom corner is a close-up of the dragon's front claw. Note, both close-ups use different camera positions than the main view.}
  \label{fig:dragon}
  
\end{figure*}

\section{Results and Evaluation}
\label{s:eval}
For the simulation, we use GPU-based open-source SPH framework openMaelstrom~\cite{winchenbach2018framework}. The radius of the fluid and boundary particles is set to $0.05$~m which corresponds to approximately $0.1$~m smoothing length. The constant $\gamma$ for the density of the boundary particles is set to $0.6$. For stabilization of  fluid-fluid interaction, we use the artificial viscosity by Monaghan~\cite{monaghan2005smoothed}, with viscosity coefficient $\alpha=0.05$. We, furthermore, use the surface tension model by Akinci~\cite{akinci2013versatile}, with a tension coefficient of $\kappa=0.15$. for the discretization and simulation, we use Cubic spline kernel~\cite{monaghan2005smoothed} and Spiky kernel gradient~\cite{mueller2004point} for values and gradients, respectively. The gravity is set to $(0, 0, -9.8)^\top~m/s^2$. 
In the experiments with resting fluids in Secs.~\ref{s:eval.sheet-no_mo}, \ref{s:eval.bulk-no_mo}, and \ref{s:eval.jet-grid} we use regular rectangular samplings with varying spacing between boundary particles, while in experiments with fluid dynamics in Secs.~\ref{s:eval.sheet-motion}, \ref{s:eval.cyl-motion}, and \ref{s:eval.dragon-dam}, we randomly sample the boundary and optimize the boundary particle positions using the method from Bell\etal\citeyear{bell2005particle} (with parameters: $100$ impulse iterations, $0.01$ impulse scale, $2$ density). 
The rest density of the fluid is set to $1,000$~$kg/m^3$. We use a fixed time step of $1$~ms, which satisfies the CFL condition in all performed experiments. As the stopping limit for the fluid solvers, we use an average density error of $0.0001\%$. For evaluation, we implemented the pressure mirroring (PM)~\cite{akinci2012versatile} and the pressure boundary (PB)~\cite{band2018pressure} methods, and we combined them with the IISPH solver in Band's modification~\cite{band2018pressure}.
To initialize the experiments, we relax the fluid using the relaxation stopping criterion by Akhunov\etal\citeyear{akhunov2023evaluation}.
All evaluations have been performed on an AMD Ryzen Threadripper 3970X CPU with $64$~GB of RAM and an NVIDIA GeForce RTX 4090 GPU with $24$~GB of VRAM. 

In all experiments, we denote the standard and decoupled versions of PB and PM as PBS, PBD, PMS and PMD, respectively.

\subsection{Resting Fluid Sheet}
\label{s:eval.sheet-no_mo}
In this experiment, we compare standard and decoupled versions of the PB and PM boundary handling methods, where the fluid sheet stays relaxed. The fluid sheet consists of $508$ fluid particles lying on the regularly sampled boundary surface. The results of the experiment are presented in Fig.~\ref{fluid_sheet_relax} and Tab.~\ref{tab:iter}.
We additionally vary the sampling distance between the boundary particles to investigate its influence on the particle height and stability. We use the term ``sampling distance'' as a ratio between the absolute distance between the neighboring boundary particles and the boundary particle diameter. We choose the sampling distances of 0.7, 0.6, and 0.5. For each distance, we measure the average height of the fluid over time and the average amplitude of velocity over time.

As can be seen from Fig.~\ref{fluid_sheet_relax}, the varying sampling distance changes both fluid height and fluid velocity for all methods. The larger the distance between boundary particles, the lower the height of the fluid. The decoupled versions of the methods, however, show a consistent height value for the PM and PB methods, while standard versions differ from each other significantly. Consequently, the standard coupled versions lead to different fluid heights and, hence, different fluid appearances and behavior. Moreover, the decoupled versions of the methods exhibit a noticeable reduction of the residual oscillations in average velocity for medium and coarser sampling distances.  

The cost for less noise and consistent fluid-boundary distance is the number of iterations; see  Tab.~\ref{tab:iter}. For all sampling distances, the decoupled versions require more iterations than the standard, coupled versions, although the difference between the standard PM and the decoupled one is minimal. 

\begin{table}
\centering
  \caption{No motion fluid sheet experiment. Number of solver's iterations}
  \label{tab:iter}
  \begin{tabular}{c|c|c|c|c}
    Sampling dist.&PBS&PBD&PMS&PMD\\\hline
    0.5 & 5 & 8 & 4 & 5\\
    0.6 & 5 & 8 & 4 & 5\\
    0.7 & 5 & 7 & 4 & 5\\
\end{tabular}
\end{table}

\subsection{Resting Fluid Bulk}
\label{s:eval.bulk-no_mo}
This experiment is similar to the prior experiment, but instead of the sheet, a bulk fluid in a container is simulated. The fluid bulk consists of $4,684$ fluid particles staying in the regularly sampled rigid container of sizes $2\times2\times8$~m. 
The results of the experiment are presented in Fig.~\ref{still_water} and Tab.~\ref{tab:iter2}, where the 2D histograms show the color-coded pressure and velocity distributions over time proposed by~\cite{akhunov2023evaluation}.

The 2D histograms indicate a significantly reduced level of noise for the decoupled versions of the boundary handling methods (see Fig.~\ref{still_water}). There is a significant decrease in the residual pressure and the residual velocity for both, particles at the boundary and in bulk. There is a significant difference in both, the pressure and the velocity distributions for PM, whose standard version is known to be very noisy~\cite{band2018pressure}. The decoupled version of PM is less noisy compared to the standard one, and even less noisy than the decoupled version of PB.

Fig.~\ref{fig:teaser} visualizes the residual velocity amplitude for the fluid bulk experiment. It can be observed that the decoupling significantly reduces the residual velocity of the fluid particles close to the boundary and the free surface of the fluid.

As for the fluid sheet experiment in Sec.~\ref{s:eval.sheet-no_mo}, the number of iterations for this experiment is higher for the decoupled versions, as depicted in Tab.~\ref{tab:iter2}. For PM, the difference in iterations is small, similar to the prior experiment.

\begin{table}
\centering
\caption{No motion bulk experiment. Number of solver's iterations}
  \label{tab:iter2}
  \begin{tabular}{c|c|c|c}
    PBS&PBD&PMS&PMD\\\hline
    5 & 7 & 4 & 5\\
\end{tabular}
\end{table}

\subsection{Fluid Sheet with Tangential Motion}
\label{s:eval.sheet-motion}
In the next experiment, we compare the standard and the decoupled methods for a fluid sheet in tangential motion over the boundary surface. We sample the boundary surface randomly and then optimize the location of the boundary particles using the method by Bell\etal\citeyear{bell2005particle}. The fluid sheet is relaxed according to Akhunov\etal\citeyear{akhunov2023evaluation}, then accelerated in x-direction up to $1$~$m/s$, and, finally, it slides freely without external acceleration besides gravity. In Fig.~\ref{fluid_sheet_slide}, we present the first $10$~s of the sliding sheet. Similar to the experiment of the resting fluid sheet in Sec.~\ref{s:eval.sheet-no_mo}, the fluid height depends on the boundary handling method in the case of the standard version, while the height of decoupled versions is very similar. Over the time of $10$~s, the average velocity decreases more quickly for the decoupled versions than for standard ones. This happens because of the reduced height of the fluid for the decoupled versions, \ie, the fluid lies closer to the boundary, and, hence, the boundary particle ``features'' cause more collisions with the fluid particles. These collisions dissipate the kinetic energy of the fluid particles. 

\subsection{Released Cylinder}
\label{s:eval.cyl-motion}
Similar to the previous experiment in Sec.~\ref{s:eval.sheet-motion}, this experiment investigates the influence of the decoupling on a fluid's tangential motion. In this case, a fluid bulk is in a cylindrical container and after relaxation it is released, yielding more complex motion than in a single fluid sheet. 
The ground surface is sampled randomly and optimized with the method by Bell\etal\citeyear{bell2005particle}. The number of fluid particles is $103,166$. 
Due to the varying fluid-boundary distance, the height of the relaxed fluid depends on the boundary handling method and on the coupling/decoupling scheme applied; see Tab.~\ref{tab:heights}, top row. Therefore, we enlarged the initial cylinder of $1$~m radius to $3$~m to achieve similar fluid heights, which leads to comparable potential energy in the system. In this case, the difference between heights is small, as it is shown in Tab.~\ref{tab:heights}, bottom row.

Fig.~\ref{cylinder} depicts the top view of the scene $1$~s after the fluid was released. It can be observed, that the advancing fluid front  reaches slightly larger distances for the standard versions of the boundary handling methods than the distance reached by the decoupled version. This is due to the higher number of collisions with the boundary ``features'' as described in Sec.~\ref{s:eval.sheet-motion}. Moreover, the advancing front of the fluid is slightly more jagged for the decoupled version, which reflects the more accurate interaction of the fluid with the boundary surface structure.

\begin{table}
\centering
  \caption{Cylinder experiment. The height of the relaxed fluid bulk in a cylinder with a radius of $1$~m and $3$~m.}
  \label{tab:heights}
  \begin{tabular}{c|c|c|c|c}
    Height&PBS&PBD&PMS&PMD\\\hline
    $1$~m & 16.45 & 14.48 & 15.98 & 14.58\\
    $3$~m & 12.919 & 12.986 & 12.933 & 12.986\\
\end{tabular}
\end{table}

\subsection{Water Jet Passing a Grid}
\label{s:eval.jet-grid}
In this experiment, a water jet is ejected on a grid with regular quadratic holes. The size of the holes is $2\times2$ particle diameter, and the rod width between the holes is also $2$ particle diameter. The results are presented in Fig.~\ref{fig:container_holes2}.

The results show that both standard coupled approaches, PMS and PBS, allow only a limited amount of water to pass through the grid, while the most fluid volume gets blocked and gathers on top of the grid. The decoupled versions, PMD and PBD, let the whole amount of liquid pass the grid and no fluid gathers on the grid. 
This different behavior results from the erroneous coupled density calculation, yielding an overestimated density force in the vicinity of fine boundary structures such as grid holes. Our decoupled density estimation method results in a proper passing of the water jet through the grid due to the consistent fluid-boundary distance. 

\subsection{Dragon Dam Break}
\label{s:eval.dragon-dam}
The last experiment consists of a fluid volume of $745$k fluid particles hitting a rigid dragon. The dragon is randomly sampled with $47$k boundary particles and optimized with the method by Bell\etal\citeyear{bell2005particle}. The results of the experiment are presented in Fig.~\ref{fig:dragon}.

The main difference between the coupled and decoupled methods in Fig.~\ref{fig:dragon} is that the coupled methods show much more splashy fluid behavior, which can be explained by unintended fluid motion near the boundary discussed in the ``Resting Fluid Bulk'' experiment in Sec.~\ref{s:eval.bulk-no_mo}. The overestimation of the fluid-boundary forces leads to the overestimation of fluid particles' velocity in the vicinity of the boundary and as a result brings an excess fluid motion, which in this case is a splashy fluid flow. 

Some more subtle differences between the coupled and the decoupled method get visible around the fine details of the dragon. Some examples are shown in the insets in Fig.~\ref{fig:dragon}, showing close-ups of the dragon's tail (bottom left) and dragon's front claw (bottom right). These close-ups show that the decoupled density estimation yields a fluid flow that is significantly closer to the dragon in small cavities and that the flow bypasses the dragon's volume more smoothly.

\section{Discussion and Limitations}
\label{s:discussion}

\textbf{Generalizability.} While we demonstrated our decoupled boundary handling for particle-based boundary representations only, our decoupling method can be applied to any boundary handling approach which is based on density estimations.
In the case of boundary integrals, for example, the iteration over the boundary particles for the density estimation can be replaced by the boundary integral calculation. Based on this density, the boundary pressure can be calculated using the boundary integral. In principle, the decoupling algorithm remains unchanged.

\textbf{Performance.} Tab.\ref{tab:iter} states that our method demands more iteration to converge than the regular methods. For the PB method, the difference is quite significant, however, for the PM method the difference is tolerable and only mildly affects the overall simulation time and performance. Depending on the application, our method may demand a trade-off between computation time and consistent fluid behavior. 

\textbf{Convergence.} Although our method converges to the same value of the average density error, the residual fluid velocity in the vicinity of the boundary is much smaller, which makes the fluid behavior much more consistent between bulk regions and the regions near the boundary. 

\textbf{Fluid-boundary coupling.} So far, our method supports only one-way coupling between fluid and boundary, and further investigations are required to combine our decoupling approach with methods like the Akinci\etal\citeyear{akinci2012versatile} or Gissler\etal\citeyear{gissler2019interlinked} to achieve two-way coupling.

\section{Conclusion}
\label{s:conclusion}
In this paper, we propose a new method that decouples forces induced by fluid-fluid interactions and fluid-boundary interactions for SPH-based fluid simulations. The decoupling is realized by separately estimating the density and calculating pressure for fluid-fluid and fluid-boundary interactions. This method prevents the overestimation of fluid density near the boundary and prevents too large distances between the fluid particles close to the boundary and between fluid and boundary particles. 
Preventing density overestimation makes the fluid flow more accurate and consistent with the bulk flow close to the boundary and at narrow and detailed boundary features. For example, the flow through small holes can be simulated accurately, while for standard methods small holes become almost impenetrable. 
Moreover, the decoupling reduces the amplitude of residual pressure and velocity, and it yields fewer oscillations at free surfaces.
As a drawback, our method requires a larger number of iterations to converge and requires more investigation toward two-way coupling between fluid and boundary.

\paragraph{Data Availability}
The datasets generated during and/or analysed during the current study are available from the corresponding author on reasonable request.


\bibliographystyle{ACM-Reference-Format}
\bibliography{bibfile}

\end{document}